\documentclass[12pt]{article}
\usepackage{amssymb}
\usepackage{epsfig}
\usepackage{float}
\usepackage{graphicx}
\usepackage{amsmath}
\newcommand{\nua}[1]{\ensuremath{\rlap
           {\kern-2.5pt\ensuremath
           {\overset{\scriptscriptstyle(-)}{\phantom{\nu}}}}
           {\ensuremath{{\nu}_{#1}}}}}

\begin{document}

\begin{center}
{\bf On atmospheric neutrino mass-squared difference in the precision era}
\end{center}

\begin{center}
S.  Bilenky
\end{center}
\begin{center}
{\em  Joint Institute for Nuclear Research, Dubna, R-141980,
Russia\\}
{\em TRIUMF
4004 Wesbrook Mall,
Vancouver BC, V6T 2A3
Canada\\}
\end{center}
\begin{abstract}
With the measurement of the small parameter $sin^{2}\theta_13$ experiments
on the study of neutrino oscillations enter into a high precision era.
I discuss here the problem of the definition of the atmospheric mass-
squared difference which will be important for analysis of data of future experiments.
\end{abstract}

Discovery of neutrino oscillations driven by small neutrino mass-squared differences and neutrino mixing is one of the most important recent discovery in the particle physics. Small neutrino masses is an evidence of a new scale in physics, presumably much larger that the electroweak scale.

First neutrino oscillation data were interpreted as  the two-neutrino $\nu_{\mu}\rightleftarrows \nu_{\tau}$
oscillations in the atmospheric range of $L/E$ and $\bar\nu_{e}\rightleftarrows \bar \nu_{\mu,\tau}$ oscillations in the solar (KamLAND) range of $L/E$ (see \cite{Bilenky:1998dt}). These oscillations were described by four parameters: $\Delta m_{23}^{2}$
and $\sin^{2}2\theta_{23}$ (the atmospheric and accelerator long baseline  experiments) and
$\Delta m_{12}^{2}$ and $\sin^{2}2\theta_{12}$ (the reactor KamLAND experiment).

With measurement of the small parameter $\sin^{2}2\theta_{13}$ in the T2K \cite{Abe:2015awa}, Daya Bay \cite{An:2015rpe}, RENO \cite{RENO:2015ksa} and Double CHOOZE \cite{Abe:2015rcp} experiments the study of neutrino oscillations enter into a new era, the era of the high precision  measurements. At this stage  a \% effects of the three-neutrino mixing are planned to be revealed.

In the case of the three-neutrino mixing we have
\begin{equation}\label{Mix}
    \nu_{lL}=\sum^{3}_{i=1}U_{li} \nu_{iL}.\quad l=e,\mu,\tau
\end{equation}
Here $\nu_{lL}$ is the flavor neutrino field, $\nu_{i}$ is the field of neutrino  with mass $m_{i}$. The the unitary PMNS mixing matrix $U$ is characterized by three mixing angles and one $CP$ phase and in the standard parametrization has the form
\begin{eqnarray}
U=\left(\begin{array}{ccc}c_{13}c_{12}&c_{13}s_{12}&s_{13}e^{-i\delta}\\
-c_{23}s_{12}-s_{23}c_{12}s_{13}e^{i\delta}&
c_{23}c_{12}-s_{23}s_{12}s_{13}e^{i\delta}&c_{13}s_{23}\\
s_{23}s_{12}-c_{23}c_{12}s_{13}e^{i\delta}&
-s_{23}c_{12}-c_{23}s_{12}s_{13}e^{i\delta}&c_{13}c_{23}
\end{array}\right),
\label{unitmixU1}
\end{eqnarray}
where $c_{12}=\cos\theta_{12}$,  $s_{12}=\sin\theta_{12}$ etc.

 Usually, in accordance with the solar neutrino data, neutrino masses are labeled in such a way that $m_{2}>m_{1}$ and $\Delta m_{12}^{2}\equiv\Delta m_{S}^{2}>0$
 is the solar mass-squared difference.\footnote{The mass-squared difference $\Delta m_{ki}^{2}$ is determined as follows $\Delta m_{ki}^{2}=m_{i}^{2}-m_{k}^{2}$.} In the case of the three neutrinos two neutrino mass spectra are possible:
\begin{enumerate}
  \item Normal spectrum (NS) : $\Delta m_{12}^{2}$ is the  difference between square of masses of the lightest neutrinos:
 $m_{3}>m_{2}>m_{1}$
  \item Inverted spectrum (IS): $\Delta m_{12}^{2}$ is the  difference between square of masses of the heaviest neutrinos: $m_{3}<m_{1}<m_{2}$.
  \end{enumerate}
Accuracies of the existing neutrino oscillation data do not allow to establish the character
 of the neutrino mass spectrum. It is one of the major problem of the future high precision neutrino oscillation experiments.

In the case of the three-neutrino mixing   neutrino transition probabilities depend on sixth  oscillation parameters. In all analysis of the neutrino oscillation data parameters $\theta_{12}, \theta_{23},\theta_{13}$, $\delta$  and  $\Delta m_{S}^{2}$, which are determined in the same way for the normal and inverted neutrino mass spectra, are used. We will discuss now the problem of the choice of atmospheric mass-squared difference, the sixths neutrino oscillation parameter.

In literature exist  {\em different definitions of  the atmospheric neutrino mass-squared difference.}
\begin{enumerate}
  \item In \cite{Bilenky:2015xwa,Abe:2015awa} this parameter is determined as a modulus of  a difference of square of the mass of $\nu_{3}$ and
square of the mass of the "intermediate" neutrino ($\nu_{2}$ in NS case and $\nu_{1}$ in IS case):
\begin{equation}\label{Amass}
\Delta m_{A}^{2}=\Delta m_{23}^{2} ~(NS),\quad  \Delta m_{A}^{2}=|\Delta m_{13}^{2}|~ (IS).
\end{equation}
\item 
The Bari group \cite{Capozzi:2015uma} determines the atmospheric mass-squared difference as follows
  \begin{equation}\label{Bari}
(\Delta m_{A}^{2})_{B}= \frac{1}{2}|\Delta m_{13}^{2}+\Delta m_{23}^{2}| ~(NS,IS)
\end{equation}  
  \item The NuFit group \cite{Gonzalez-Garcia:2014bfa} determines the atmospheric mass-squared difference in the following way
 \begin{equation}\label{NuFit}
 (\Delta m_{A}^{2})_{NF}=\Delta m_{13}^{2}~(NS)~~ (\Delta m_{A}^{2})_{NF}= |\Delta m_{23}^{2}|~ (IS)
    \end{equation}
\item In analysis of the data of Daya  Bay \cite{An:2015rpe} and RENO \cite{RENO:2015ksa}  experiments the following "large" mass-squared difference was used
\begin{equation}\label{Daya}
\Delta m^{2}_{ee}=\cos^{2}\theta_{12}\Delta m^{2}_{13}+\sin^{2}\theta_{12}\Delta m^{2}_{23}
\end{equation}
\end{enumerate}
Let us notice that with the definition given in 1. vacuum neutrino transition probabilities have the  simple form of the sum of atmospheric, solar and interference terms \cite{Bilenky:2015xwa}
\begin{eqnarray}
&&P^{NS}(\nua{l}\to \nua{l'})
=\delta_{l' l }
-4|U_{l 3}|^{2}(\delta_{l' l} - |U_{l' 3}|^{2})\sin^{2}\Delta_{A}\nonumber\\&&-4|U_{l 1}|^{2}(\delta_{l' l} - |U_{l' 1}|^{2})\sin^{2}\Delta_{S}
-8~[\mathrm{Re}~(U_{l' 3}U^{*}_{l 3}U^{*}_{l'
1}U_{l 1})\cos(\Delta_{A}+\Delta_{S})\nonumber\\
&&\pm ~\mathrm{Im}~(U_{l' 3}U^{*}_{l 3}U^{*}_{l'
1}U_{l 1})\sin(\Delta_{A}+\Delta_{S})]\sin\Delta_{A}\sin\Delta_{S},
\label{Genexp5}
\end{eqnarray}
and
\begin{eqnarray}
&&P^{IS}(\nua{l}\to \nua{l'})
=\delta_{l' l }
-4|U_{l 3}|^{2}(\delta_{l' l } - |U_{l' 3}|^{2})\sin^{2}\Delta_{A}\nonumber\\&&-4|U_{l 2}|^{2}(\delta_{l' l} - |U_{l' 2}|^{2})\sin^{2}\Delta_{S}
-8~[\mathrm{Re}~(U_{l' 3}U^{*}_{l 3}U^{*}_{l'
2}U_{l 2})\cos(\Delta_{A}+\Delta_{S})\nonumber\\
&&\mp ~\mathrm{Im}~(U_{l' 3}U^{*}_{l 3}U^{*}_{l'
2}U_{l 2})\sin(\Delta_{A}+\Delta_{S})]\sin\Delta_{A}\sin\Delta_{S}.
\label{Genexp6}
\end{eqnarray}
Here $\Delta_{A,S}=\frac{\Delta m^{2}_{A,S}L}{4E}$, where $L$ is the detector-source distance and $E$ is the neutrino energy.

It is obvious that parameters $\Delta m_{A}^{2}, (\Delta m_{A}^{2})_{B}, (\Delta m_{A}^{2})_{NF}$
are determined in the same way for normal and inverted spectra. We  have
\begin{equation}\label{Connect}
(\Delta m_{A}^{2})_{B}=\Delta m_{A}^{2}+\frac{1}{2}\Delta m_{S}^{2}, \quad  (\Delta m_{A}^{2})_{NF}=\Delta m_{A}^{2}+\Delta m_{S}^{2}.
\end{equation}
From analysis of neutrino oscillation data it follows that $\frac{\Delta m_{S}^{2}}{\Delta m_{A}^{2}}\simeq 3\cdot 10^{-2}$. Thus, different definitions of the atmospheric mass-squared difference differ by a few \%. However, the goal of future neutrino oscillation experiments is to measure oscillation parameters with a \% accuracy. In the precision era {\em one definition of atmospheric mass-squared difference will be definitely important.} Theoretically there is no preferred definition. From our point of view a consensus must be found what definition is the most suitable from the practical point of view.

The "averaged mass-squared difference" \begin{equation}\label{Daya1}
\Delta m^{2}_{ll}=\cos^{2}\theta_{ll}\Delta m^{2}_{13}+\sin^{2}\theta_{ll}\Delta m^{2}_{23},\quad l=e,\mu
\end{equation}
was introduced in \cite{Nunokawa:2005nx}. Here
\begin{equation}\label{Mix6}
\cos^{2}\theta_{ll}=    \frac{|U_{l1}|^{2}}
{|U_{\l1}|^{2}+|U_{\l2}|^{2}},\quad   \sin^{2}\theta_{ll}=  \frac{|U_{l2}|^{2}}
{|U_{l1}|^{2}+|U_{l2}|^{2}}.
\end{equation}
The probability of $\nua{l}$ to survive in vacuum can be presented in the  form
\begin{equation}\label{Mix4}
P(\nua{l}\to\nua{l})=1-4|U_{\l 1}|^{2}|U_{\l 2}|^{2}\sin^{2}\Delta_{12}-4|U_{l3}|^{2}(1-|U_{l3})|^{2}(\cos^{2}\theta_{ll}\sin^{2}\Delta _{13}+\sin^{2}\theta_{ll}\sin^{2}\Delta_{23}),
\end{equation}
where  $\Delta_{ki}=\frac{\Delta m^{2}_{ki}L}{4E}$. We have $\Delta m^{2}_{13}=\Delta m^{2}_{23}+\Delta m^{2}_{12}$. Taking into account this relation from (\ref{Daya1}) we find
\begin{equation}\label{Mix8}
\Delta m^{2}_{13}=\Delta m^{2}_{ll}+\sin^{2}\theta_{ll}\Delta m^{2}_{12},\quad \Delta m^{2}_{23}=\Delta m^{2}_{ll}-\cos^{2}\theta_{ll}\Delta m^{2}_{12}.
\end{equation}
In reactor and  long baseline accelerator experiments $\Delta_{12}\ll 1$ and  $|\Delta_{13}|\simeq
|\Delta_{23}|\simeq 1$. It is easy to see that in the expansion of the term $(\cos^{2}\theta_{ll}\sin^{2}\Delta _{13}+\sin^{2}\theta_{ll}\sin^{2}\Delta_{23})$ over $\Delta_{12}$ the liner  term disappear. If we neglect the quadratic term, for $\bar\nu_{e}\to\bar\nu_{e}$ transition probability from (\ref{Mix4}) we find
\begin{equation}\label{Mix11}
P(\bar\nu_{e}\to\bar\nu_{e})\simeq 1-\cos^{4}\theta_{13}\sin^{2}2\theta_{12}\sin^{2}\Delta_{S}
-\sin^{2}2\theta_{13}\sin^{2}\Delta_{ee}.
\end{equation}
This expression was used for the analysis of the latest  Daya Bay \cite{An:2015rpe} and RENO \cite{RENO:2015ksa} data. We would like now to comment the usage of the parameters $\Delta m^{2}_{ee}$ and $\Delta m^{2}_{\mu\mu}$.
\begin{itemize}
  \item These parameters describe  data of only disappearance experiments.
 \item Their definition depends on the character of neutrino mass spectrum. In fact, we have
 \begin{equation}\label{Mix13.}
 \Delta m^{2}_{ee}=\Delta m^{2}_{A}+\cos^{2}\theta_{12}\Delta m^{2}_{S}~(NS),~~
 | \Delta m^{2}_{ee}|=\Delta m^{2}_{A}+\sin^{2}\theta_{12}\Delta m^{2}_{S}~(IS).
 \end{equation}
\item
In order to determine the fundamental parameter
$\Delta m^{2}_{A}$ from the measured value of $| \Delta m^{2}_{ee}|$ and compare the reactor, atmospheric and accelerator data  we need to know $\sin^{2}\theta_{12}$, $\Delta m^{2}_{S}$ and also the neutrino mass spectrum.
\end{itemize}
Convenient alternative expressions the $\bar\nu_{e}$ survival probability in the case of the normal and inverted neutrino mass spectra, which follow from (\ref{Genexp5}) and (\ref{Genexp6}),  have the form
\begin{eqnarray}
&&P^{\mathrm{NS}}(\bar\nu_{e}\to \bar\nu_{e})=1-
\sin^{2}2\theta_{13} \sin^{2}\Delta_{A}
\nonumber\\
&&-
(\cos^{4}\theta_{13}\sin^{2}2\theta_{12}+\cos^{2}
\theta_{12}\sin^{2}2\theta_{13}) ~ \sin^{2}\Delta_{S}
\nonumber\\
&&-2\sin^{2}2\theta_{13}\cos^{2}\theta_{12} ~\cos(\Delta_{A}+\Delta_{S}) \sin\Delta_{A}\sin\Delta_{S}.\label{3nue4}
\end{eqnarray}
and
\begin{eqnarray}
&&P^{\mathrm{IS}}(\bar\nu_{e}\to \bar\nu_{e})=1-
\sin^{2}2\theta_{13} \sin^{2}\Delta_{A}
\nonumber\\
&&-
(\cos^{4}\theta_{13}\sin^{2}2\theta_{12}+\sin^{2}
\theta_{12}\sin^{2}2\theta_{13}) ~ \sin^{2}\Delta_{S}
\nonumber\\
&&-2\sin^{2}2\theta_{13}\sin^{2}\theta_{12} ~\cos(\Delta_{A}+\Delta_{S})\sin\Delta_{A}\sin\Delta_{S}.\label{3nue5}
\end{eqnarray}
We believe that in the era  of the high precision  neutrino oscillation experiments, {\em data must be analyzed in terms of universal fundamental neutrino oscillation parameters} (mixing angles, phase and  independent mass-squared differences) which characterize all transition probabilities and are directly connected with neutrino mixing matrix and masses.

I would like to thank C. Giunti for the interesting discussion.

\end{document}